\documentclass[useAMS]{mn2e}

\def\cm3{\hbox{cm$^{-3}$}}

\usepackage{psfig}
\usepackage{graphicx}
\title[Gas expulsion and the destruction of massive young clusters]
{Gas expulsion and the destruction of massive young clusters }
\author[S.P.~Goodwin and N.~Bastian] {Simon~P.~Goodwin$^1$ and Nate~Bastian$^2$ \\
$^1$Department of Physics and Astronomy, The University of Sheffield,
  Hicks Building, Hounsfield Road, Sheffield S3 7RH\\
$^2$Department of Physics and
Astronomy, University College London, Gower Street, London, WC1E 6BT\\}
\date{Accepted. Received; in original form}
\pagerange{\pageref{firstpage}--\pageref{lastpage}}
\pubyear{2005}
\begin{document}
\maketitle
\label{firstpage}
\begin{abstract}
We examine the luminosity and dynamical mass estimates for young
massive stellar clusters.  For many young ($<50$~Myr) clusters, the
luminosity and dynamical mass estimates differ by a significant
amount.  We explain this as being due to many young clusters being out
of virial equilibrium (which is assumed in dynamical mass estimates)
because the clusters are undergoing violent relaxation after expelling
gas not used in star formation.  We show that, if we assume that
luminous mass estimates are correct (for a standard IMF), at least
50~per cent of young clusters for which dynamical masses are
known are likely to  be destroyed within a few 10s~Myr of their
formation.  Even clusters which will retain a bound core may lose a
large fraction of their stellar mass.  We also show that  the core
radius and other structural parameters change significantly during the
violent relaxation that follows gas expulsion and that they should be
considered instantaneous values only, not necessarily reflecting the
final state of the cluster. In particular we note that the increasing
core radii observed in young LMC/SMC clusters can be well explained as
an effect of rapid gas loss.
\end{abstract}

\begin{keywords} galaxies: star clusters -- stellar dynamics --
  methods: $N$-body simulations
\end{keywords}

\section{Introduction}\label{intro}

Almost all stars form in star clusters which have a mass function of
the form $N(M)  \propto M^{-2}$ (Lada \& Lada 2003).  This means that
most clusters are fairly small ($10^2$ -- $10^3 M_\odot$), however
some of the most interesting young clusters are the rare, massive
`super star clusters' (SSCs) with masses comparable to, or greater
than, the Galactic globular clusters.  SSCs are found in merging and
interacting galaxies such as NGC~7252 (Miller et al.~1997) and
NGC~6745  (de Grijs et al.~2003)\footnote{See Larsen~(2004) for an
extensive list of starburst, merger, spiral, dwarf, and irregular
galaxies which contain populations of young massive clusters.}, as
well as the LMC (Elson, Fall \& Freeman 1989; Elson 1991; Mackay \&
Gilmore 2003), and even within our own Galaxy (esp. Wd~1, see Clark et
al. 2005).

Star clusters form within giant molecular clouds and remain embedded
in their natal clouds for $\sim 2$ -- $5$~Myr before the combined effect of
massive stellar winds and supernovae drive-out the gas not used in
star formation.  This gas represents $\sim 50$~\% of the initial
mass of the cluster and its rapid loss (on less than a crossing 
time, Goodwin 1997a; Melioli \& de Gouveia dal Pino 2006) causes 
a huge change to the potential of the
cluster.  The stars left behind after gas expulsion violently relax to
the new potential and attempt to return to virial equilibrium.  

The effect of gas expulsion on the (stellar) dynamics of star clusters 
is well understood theoretically (e.g. Tutkov 1978; Hills 
1980; Goodwin 1997a,b; Boily 
\& Kroupa 2003a,b).  The stars have too large a velocity dispersion
for their new potential and the cluster expands, and stars with a
velocity greater than the new escape velocity are lost.  The loss of
stars is not instantaneous, however, and escaping stars remain physically
associated with the cluster for 10s~Myr.

Recently Bastian \& Goodwin (2006, hereafter BG06) have shown that a 
number of massive young clusters show the signature of gas expulsion:
an excess of light at large radii over and above a relaxed (King or 
Elson-Fall-Freeman (EFF)) surface brightness profile.  They note that 
such signatures are also seen in many young LMC clusters (see Elson 
et al. 1989; Elson 1991; Mackay \& Gilmore 2003).

BG06 show that a consequence of gas expulsion is that many young 
clusters {\em are not in virial equilibrium}: thus dynamical mass estimates
based on virial equilibrium may be highly inaccurate.  In particular,
claims of unusual IMFs based on dynamical masses being inconsistent
with luminous mass estimates are probably incorrect.

In this paper we take the approach that luminosity mass estimates are
correct and the discrepancy between luminosity and dynamical masses
can be used as an indicator of the extent to which young clusters are
out of virial equilibrium.  This may then be used to estimate the star
formation efficiency (SFE; see below for details)  of young clusters
and their survivability.  Additionally, we will show that rapid gas
removal can explain the trend of increasing core radii with age
observed in young LMC clusters.  This paper is organised as follows;
in \S~\ref{sec:sims} we introduce the simulations of rapid gas loss in
cluster and address the caveats in the current study. In
\S~\ref{sec:results} we present the results of the simulations placing
particular emphasis on the dynamical state of young clusters and on
the effects of ``infant-weight loss'' and ``infant mortality''.  We
summarise our results in \S~\ref{sec:conclusions}.

\section{Simulations}
\label{sec:sims}

We have simulated the $N$-body dynamics of a cluster post-gas
expulsion using the GRAPE-5A special-purpose hardware at the
University of Cardiff (Kawai et al. 2000) using a simple $N$-body integrator code.

\subsection{Initial conditions}

Star clusters were modelled as a Plummer (1911) sphere with a density
distribution $\rho(r)$ of the form
\begin{equation}
\rho(r) = \frac{3M_p}{4\pi R_P^3} \frac{1}{(1 + (r/R_P)^2)^{5/2}}
\end{equation}
where $M_P$ is the total mass of the cluster, and $R_P$ is the Plummer
radius (the half-mass radius is $r_{\rm half} \sim 1.3 R_P$).  The
initial positions and velocities of the particles were constructed
using the method of Aarseth, H\'enon \& Wielen (1974).  The
cluster changes rapidly after gas expulsion and so the exact form of
the initial density distribution is unimportant as the cluster loses
memory of this configuration very quickly (see e.g. Goodwin 1997a).

A number of authors have modelled the effect of gas expulsion on star
clusters (e.g. Lada, Margulis \& Dearborn 1984; Goodwin 
1997a,b; Geyer \& Burkert 2001;
Kroupa, Aarseth \& Hurley 2001; Kroupa \& Boily 2002; Boily \& Kroupa
2003a,b).  If the gas removal timescale is less than a crossing time 
(as it is expected to be, especially in massive clusters - see Goodwin 
1997a; Melioli \& de Gouveia dal Pino 2006) then it is effectively
instantaneous and the system can be modelled as one that is 
{\em initially} out of virial equilibrium (this avoids
  modelling the gas as an external potential (e.g. Goodwin 1997a,b;
  Kroupa et al. 2001).

We define an {\em effective star formation efficiency} (eSFE),
$\epsilon$ which is a measure of how far out of virial equilibrium the
cluster is {\em after} gas expulsion.  We define the virial ratio as
$Q=T/|\Omega|$, where $T$ is the kinetic energy, and $\Omega$ the
potential energy (so a system in virial equilibrium has $Q=0.5$).  

A cluster with an eSFE of $\epsilon$ initially has a velocity 
dispersion $\sqrt{1/\epsilon}$ too large to be in virial equilibrium.
For example, a $10^5 M_\odot$ GMC with a radius of $\sim 1$~pc would
have a velocity dispersion in virial equilibrium of $\sim 20$ km
s$^{-1}$.  If it formed stars at 50 per cent efficiency, {\em and} 
those stars were in virial equilibrium with 
the total potential of the stars and gas initially then, after 
gas expulsion, the virial ratio of the stars 
would be $Q=1$\footnote{The potential after gas expulsion is a factor
  of $\epsilon^2$ smaller, while the kinetic energy is a factor of
  $\epsilon$ lower, thus the virial ratio goes as $\epsilon^{-1}$}.  
It should be noted that if stars {\em do not}
form in virial equilibrium with the gas the eSFE is not a direct
measure of the star formation efficiency.  Indeed, it is possible for
clusters to have eSFEs in excess of 100 per cent.  For example, if 
the stars formed from the above GMC with a velocity dispersion of only 10 km
s$^{-1}$ from the GMC (thus less than the 15 km s$^{-1}$ required 
to be in virial equilibrium after gas expulsion), the eSFE would be
225 per cent and the cluster would contract after gas expulsion (in
fact it would be contracting before gas expulsion).  However, as we
shall see, all the evidence points towards eSFEs being less than 100
per cent.

As canonical initial conditions we choose a cluster with $R_P = 3.5$~pc and
$M_P/\epsilon = 5 \times 10^4 M_\odot$.  The results however scale (as
per $N$-body units) as we do not include an external tidal field or
stellar evolution which would set a physical timescale.  Excluding
these effects is probably not important as clusters are so far
out of virial equilibrium due to gas expulsion that these effects will
be small perturbations on the overall behaviour of the system (see below).  Thus
we expect the results seen here to be applicable to clusters of all
sizes\footnote{This assumption will probably fail if cluster masses
  were fairly small as the tidal field may then play an important role
  (see Kroupa et al. 2001), however for massive clusters it is
  probably correct.}.

Simulations were conducted using $N=30000$ equal-mass particles.
Tests show that the results are insensitive to both $N$ and the
softening length of the gravitational interactions for any reasonable 
values.  This convergence is unsurprising as the simulations follow the
violent relaxation of the cluster to a new equilibrium, a situation in
which 2-body encounters are fairly unimportant and it is the bulk
behaviour of the potential that dominates the evolution.

The dynamical masses of star clusters models are calculated by ``observing''
the 1D velocity dispersion and calculating a mass using
\begin{equation}
M= \eta \frac{R_{\rm hl} \sigma_{\rm 1D}^2}{G}
\end{equation} 
where $R_{\rm hl}$ is the half-light radius, $\sigma_{\rm 1D}$ is the
1D (line-of-sight) velocity dispersion, $G$ the gravitational
constant, and $\eta$ is a numerical constant $\sim 10$ (see 
Fleck et al. 2006 and references therein).  Recently, Fleck et 
al. (2006) have modelled the effect on the parameter
$\eta$ used in dynamical mass estimates.  They find that $\eta$ 
  increases dramatically due to mass segregation, such that the value
typical used by observers, $\eta = 9.75$ should underestimate the true
mass of the clusters.  As shown later in  Fig.~\ref{fig:clusters} (see 
also Bastian et al.~2006) the youngest clusters have masses which
appear overestimated with respect to their luminous 
masses.  Therefore the results of Fleck et
al. (2006) seem not to apply to the young massive clusters with
dynamical mass measurements (i.e. any cluster shown 
in Fig.~\ref{fig:clusters}) with the possible exception of M82-F.

\subsection{Caveats}

\subsubsection{Stellar evolution}
These simulations do not take into account the mass lost by stellar
evolution.  In the first few tens of~Myr, clusters may lose $>10$~per cent
of their stellar mass.  Goodwin (1997a,b) did include stellar
evolutionary mass loss and found that it was a fairly minor
perturbation on the expansion (as 10~per cent of the stellar mass is only a
few~per cent of the total initial mass).  However, mass loss from stellar
evolution may play a significant role once a cluster has relaxed into
a new equilibrium as it will cause the cluster to expand further and
possibly be disrupted even if the eSFE was high enough to allow it to
survive the initial gas expulsion.

\subsubsection{Stellar mass function and mass segregation}
We also do not include a mass function.  Goodwin (1997a) included a
mass function, whilst Goodwin (1997b) used equal-mass particles.
Again, no significant difference in the results was found.  This is
because we assume that there is no energy equipartition in these young
clusters and so all stars, whatever their mass, have the same velocity
dispersion, and so have the same probability of being lost.  In
reality, young clusters do appear to be mass segregated (e.g. de 
Grijs et al. 2001a,b).  This may mean that there is some initial 
equipartition and higher mass stars
have a lower velocity dispersion.  If so, we would
expect low-mass stars to escape preferentially as their velocity
dispersion is higher.  How significant this effect is depends on the
details of mass segregation and how far down the mass function it
extends (ie. is it only the highest-mass stars that have a lower
velocity dispersion, or does velocity dispersion depend upon mass for
{\em all} masses?).  However, we feel that this is a second-order
effect as the new escape velocity of the cluster after gas expulsion
is a factor of $>4$ lower than the initial escape velocity.

\section{Results}
\label{sec:results}

Firstly we describe the effects of gas expulsion and violent
relaxation on the structure of young massive star clusters.  Then we
examine the differences between the dynamical masses and luminosities
of a large number of young clusters and how they might be explained.
Finally we discuss infant mortality and the destruction of clusters.

\subsection{The effects of gas expulsion}

After gas expulsion stars have far too large a velocity dispersion for
their new potential.  As a result, the cluster expands in an attempt
to find a new equilibrium.  Those stars with a velocity greater than
the new escape velocity tend to escape the cluster\footnote{In reality
  it is not quite this simple as, (a) stars can redistribute energy
  through 2-body encounters, and (b) the escape velocity changes as
  the cluster loses stars.  However to a first approximation this is
  what occurs.} (see also Tutkov 1978; Goodwin 1997a,b; Kroupa 
\& Boily 2002; Boily \& Kroupa 2003a,b; BG06).

\begin{figure*}
\centerline{\psfig{figure=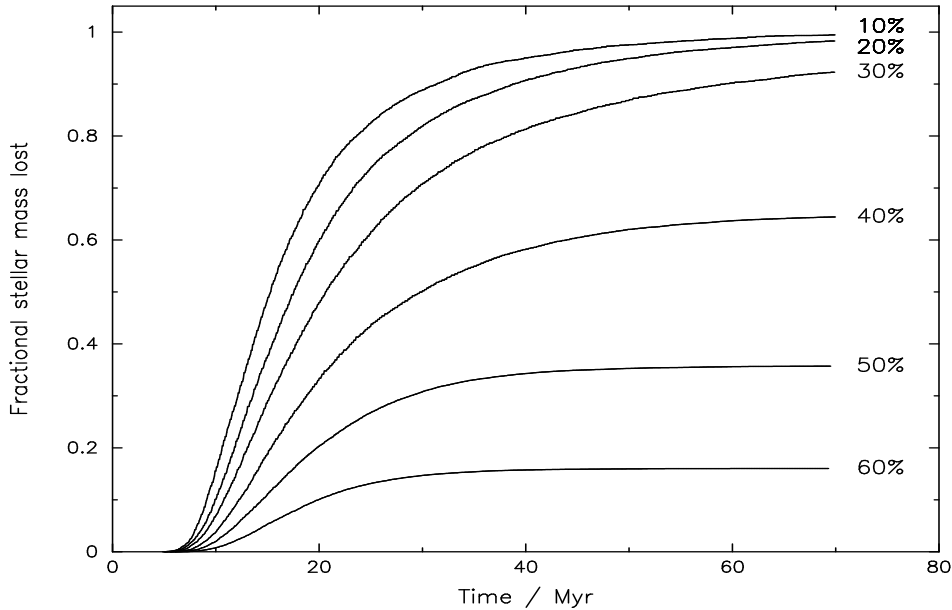,height=8.0cm,width=12.5cm,angle=270}}
\caption{The fraction of stars which are pushed over 20~pc of the centre
of mass of the cluster with time for clusters with eSFEs between 10 and
60~per cent.  Clusters with eSFEs$<30$~per cent are unbound and 
rapidly disperse into the field, but clusters with slightly 
higher eSFEs are able to retain a bound core of stars.}
\label{fig:massloss}
\end{figure*}

In Fig.~\ref{fig:massloss} we show the mass of escaped stars 
with time for different eSFEs.  If the eSFE is $< 30$~per cent then the 
cluster becomes completely unbound within a few 10s~Myr as it is 
incapable of reaching a new equilibrium.  For greater eSFEs the
cluster manages to retain a bound core but may loose a very 
significant fraction of its initial stellar mass (see also 
Goodwin 1997a,b; Kroupa \& Boily 2002; Boily \& Kroupa 2003a,b; BG06). It 
is these escaping stars that create the excess of light at large
radii that BG06 show is observed in a number of young clusters and is the
most obvious signature of violent relaxation after gas expulsion. 

\begin{figure*}
\centerline{\psfig{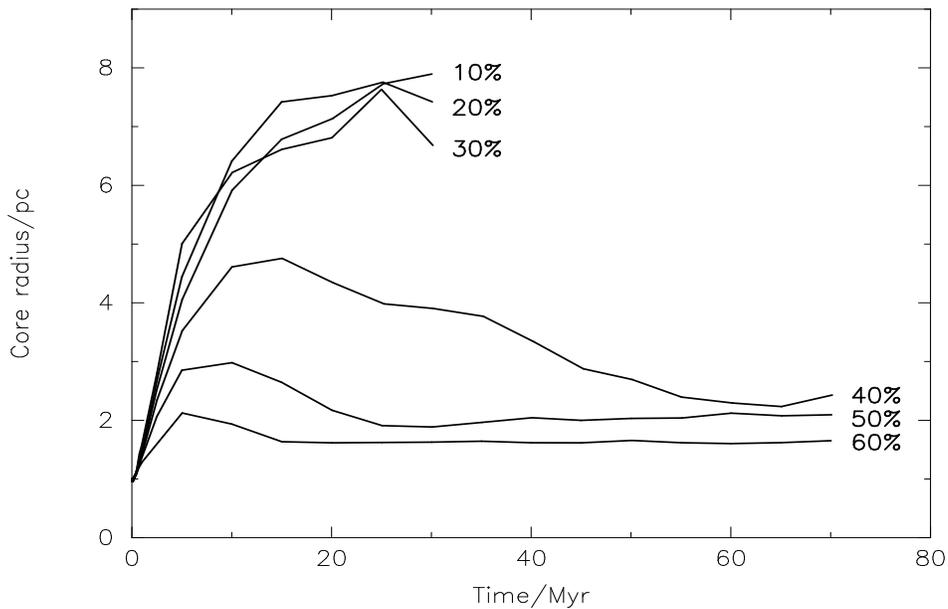}}
\caption{The evolution of the core radius as measured from the $a$ and
$\gamma$ parameters of the best-fit EFF profile (see text) with time
  for clusters with eSFEs between 10 and 60~per cent.  Note that
  measuring the core radii of low eSFE clusters at late times becomes
  very difficult due to the low numbers of stars still within a
  nominal 20~pc radius cluster, therefore we do not plot low eSFE
    clusters beyond 30 Myr as the values of the core radius become 
meaningless.}
\label{fig:coreradii}
\end{figure*}

The expansion of the clusters, in an attempt to reach a new equilibrium,
changes the core (and other) radii of clusters as illustrated in
Fig.~\ref{fig:coreradii}.  This figure shows the core radii measured
by fitting an EFF profile (Elson, Fall \& Freeman 1987) to the surface
brightness profile of the form
\begin{equation}
\mu(r) = \mu(0) (1 + r^2/a^2)^{-\gamma/2}
\end{equation}
where $\mu(0)$ is the central surface brightness, $\gamma$ is a
measure of the steepness of the outer parts of the profile, and $a$ is
related to the core radius such that $r_c \sim a(2^{2/\gamma} -
1)^{1/2}$.  Note that after 20 -- 30~Myr, for low eSFE clusters,
the lack of stars within the nominal 20~pc cluster radius makes
determinations of the core radius very noisy and effectively
meaningless.  By such an age, low eSFE clusters have a very low
surface brightness and we doubt if such objects would ever be
observable above the background.

\begin{figure*}
\centerline{\psfig{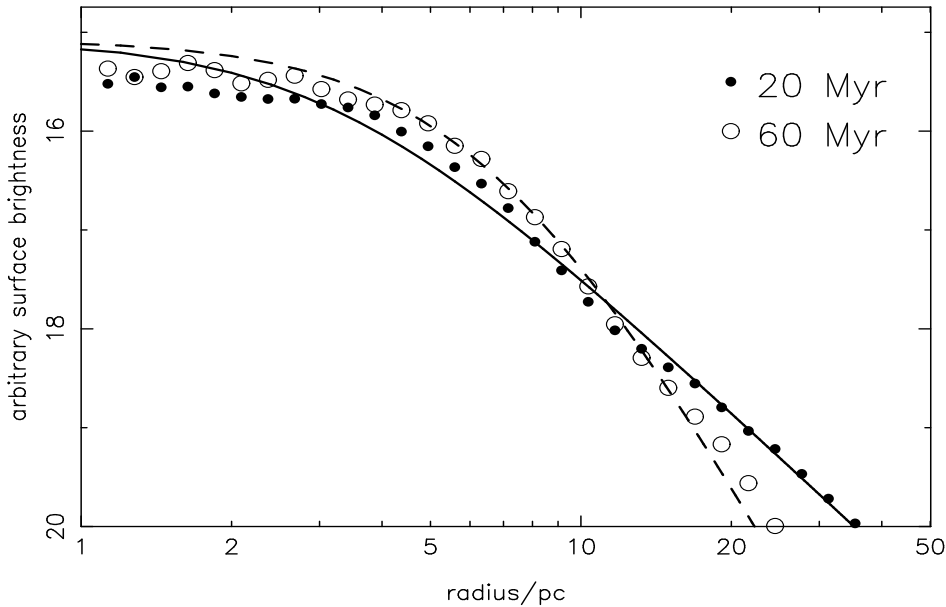}}
\caption{The surface brightness profile of a 40~per cent eSFE cluster
  at 20~Myr (filled circles) and 60~Myr (open circles) with the
  best (least squares)-fit EFF profiles for each cluster (full and
  dashed-lines respectively).  The surface brightness is scaled
  arbitrarily from the surface mass density.}
\label{fig:fitting}
\end{figure*}

Clearly, the core radii increase as the clusters expand.  However,
much of the increase in the core radii of clusters with eSFE $\sim
40$~per cent is due to fitting the EFF profile to the {\em entire}
cluster, including the tail of escaping stars.  In
Fig.~\ref{fig:fitting} we show the surface brightness profiles of a 40~per
cent eSFE cluster after 20 and 60~Myr together with their best-fit EFF 
profiles.  At 20~Myr, the best-fit EFF profile is clearly not a good
fit, this is due to attempting to fit an equilibrium model to a
non-equilibrium cluster.  The fit over-estimates the core radius as it
tries to fit the excess light at large radii with $a=3.2$~pc and
$\gamma=1.9$ giving $r_c=3.4$~pc.  By 60~Myr most of the excess light
at large radii has gone, as those stars have become physically
detached from the cluster, but the inner structure of the cluster is
very similar.  However, now the profile does not have to fit the
excess light the best-fit becomes $a=6.7$~pc and $\gamma=3.65$ giving
$r_c=2.1$~pc\footnote{There is still a small amount of excess light at
  large radii even after 60~Myr.  The attempt to fit the excess light
  in both profiles results in the over-estimate of the central surface
  brightness in both cases.}.

This clearly illustrates that, during the period that there is a
significant contribution to the luminosity from
escaping stars, the parameters obtained from profile fitting may have
little to do with any final parameter.  This may be avoided 
by {\em not} fitting profiles to the outer regions of clusters.  A 
significant improvement occurs if the fitting is to luminosity rather
than magnitude as this tends to weight the central regions more 
heavily.  Similar effects can occur with other parameters - in
particular the $\gamma$ parameter from the EFF profile (see
Goodwin 1997b).

This increase in core radii due to the expansion of clusters could
well account for the increase in the observed core radii of very young 
clusters (see Elson et al. 1989; Elson 1991, esp. Fig.~14; Mackey \&
Gilmore 2003).   Core radii generally increase with age as would be expected,
in particular core radii seem to increase dramatically in the first
few~Myr (however this is based on just one point - R136).  There is
also very mild evidence that the core radii start to level-off after
$\sim 50$~Myr as would be expected.  However it is difficult to draw
any conclusions from this data alone as we do not know what the
initial core radii were, nor if they were all the same (but see Goodwin
1997b).

The rapid evolution of the core radii illustrates an important point.  
When measuring the properties of young star clusters many
measurements are of {\em instantaneous} values which may not have a
simple connection to a `final' value (ie. the value when a new equilibrium
has been reached).  Measurements of the core radius of a cluster at
$20$~Myr may give a value that is far in excess of the core radius
that cluster will have at $50$~Myr.  Indeed, measurements of
parameters such as a characteristic radius need not give any clues as
to the eSFE or final fate of a cluster unless all clusters were
initially the same (e.g. see Goodwin 1997b for an attempt to combine
parameters to estimate the final fate of young LMC clusters).

\begin{figure*}
\centerline{\psfig{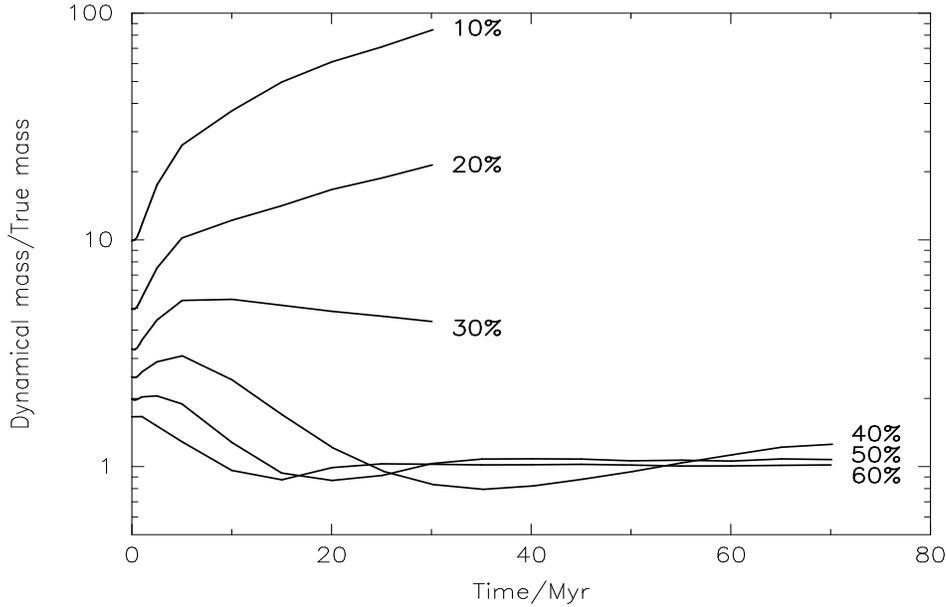}}
\caption{The ratio of dynamical mass to 'true' mass (the mass within
  20~pc of the cluster centre) against time after gas expulsion for
  eSFEs between 10 and 60 per cent.  Note that the dynamical masses 
  of disrupting (eSFEs of 10 -- 30 per cent) clusters become rather
  meaningless after $\sim 20$~Myr due to the low number of stars
  within 20~pc of the nominal cluster centre.}
\label{fig:virialtrue}
\end{figure*}

The simplest way to determine the eSFE of young clusters is
to compare the dynamical and luminous masses.  One of the main points
made by BG06 was that the dynamical masses of
clusters are not an accurate measure of their true masses during the
expansion phase.  In Fig.~\ref{fig:virialtrue} we show the evolution 
of the ratio of dynamical mass to true mass (the actual mass of stars 
within 20~pc of the cluster) for clusters with eSFEs of 10 -- 60~per cent.  
Clearly, the lower the eSFE - and so the higher the initial virial 
ratio - the worse the dynamical mass becomes as a measure of the true 
mass.

\subsection{Dynamical mass vs. luminous mass}

\begin{figure*}
\centerline{\psfig{figure=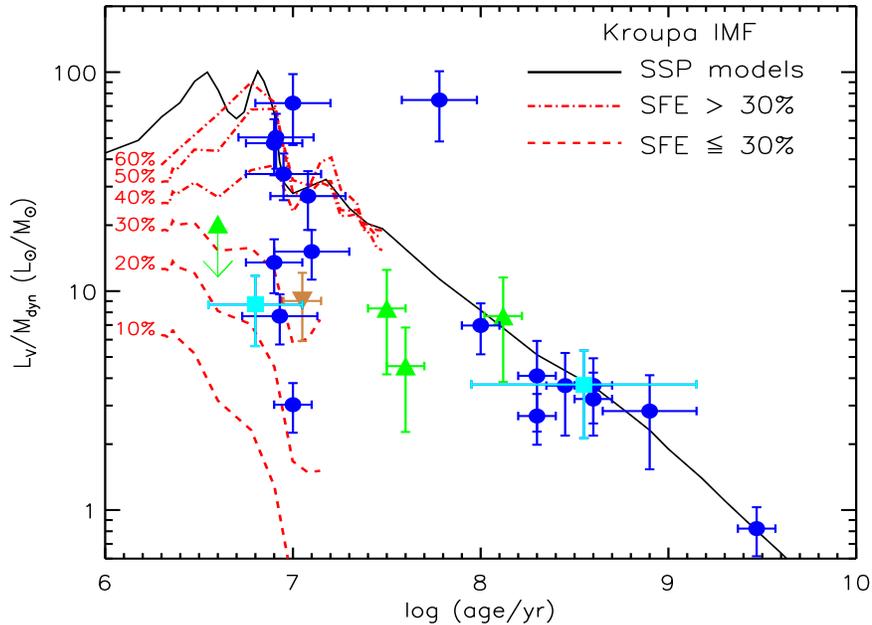,height=9cm,width=12.5cm,angle=0}}
\caption{The light-to-mass ratio of young clusters.  The circles
  (blue) are taken from Bastian et al.~2006 and 
  references therein, the triangles with errors (green) are LMC
  clusters (Mclaughlin \& van der Marel~2005), the upside down
  triangle (brown) is for 
  NGC~6946-1447 corrected for internal extinction (Larsen et al.~2006), and
  the squares (cyan) are from \"Ostlin et al.~(2006).  The triangle without errors is the tentative 
  upper limit for cluster R136 in 30~Dor (Bosch et al.~2001, Hunter et
  al.~1995). The
  solid (black) line is the prediction of simple stellar population
  models (SSPs) with a Kroupa~(2002) stellar IMF and solar metallicity from the simple
  stellar populations models of Maraston~(2005). The dashed and
  dash-dotted lines (red) are the SSP model tracks folded with the
  effects of rapid gas removal following non-100\% star-formation
  efficiencies (SFE) (ie.~Fig~\ref{fig:virialtrue}). Dashed lines
  represent the SFEs where the clusters will become completely
  unbound.  We have assumed that the residual gas has been removed
  instantly at an age of 2~Myr.} 
\label{fig:clusters}
\end{figure*}

In Fig.~\ref{fig:clusters} we show the ratio of
dynamical-mass-to-luminosity (DMtL) for a selection of clusters against 
their age (updated from Fig.~5 of Bastian et al. 2006).  The solid 
line shows the expected DMtL for virialised clusters (ie. those for
which the dynamical mass is the true mass) with a Kroupa~(2002) IMF
from the simple stellar population models of Maraston~(2005).
The luminosity of the cluster changes as it ages due to stellar
evolution which is included in the calculation of the canonical 
IMF line.  Clearly, for young clusters, the DMtL is often significantly below
this canonical value.  Older clusters, however, lie very close to this
line (Bastian et al. 2006).  The discrepancy between the expected 
and observed DMtL has been taken
as an indication of a non-standard IMF within some of these clusters.  We,
however, interpret this as a signature of violent relaxation.  

Older clusters are expected to be in virial equilibrium having survived
their gas expulsion episode.  That they lie on the standard IMF DMtL
line suggests that they have a standard IMF.  This gives us
confidence that our assumption of a standard IMF for young clusters is 
correct.  It may be that young clusters with unusual IMFs destroy 
themselves and so are not present in the old sample.  However to lie 
below the canonical line on Fig.~\ref{fig:clusters} (as all but 2
clusters do) their IMF would have to be bottom heavy (ie. less light
for their mass than expected) which would mean that they would be {\em
  more} likely to survive as the effects of stellar evolution would be 
significantly less dramatic.

Given that we expect dynamical masses to over-estimate the true mass
of a young cluster it is possible to use the luminous mass (making the
apparently reasonable assumption of a standard IMF) to determine 
how wrong the dynamical mass is, and hence the eSFE, and so final 
fate, of the cluster.  We also mark on Fig.~\ref{fig:clusters} the 
expected evolutionary paths of clusters with eSFEs between 10 and 
60~per cent, i.e. the canonical IMF SSP evolution folded with the 
results of Fig.~\ref{fig:virialtrue}.  As the dynamical mass
overestimates the true mass (see Fig.~\ref{fig:virialtrue}), the DMtL
will lie below the canonical line.  Note that the evolutionary paths
for different eSFEs start at 2~Myr to account for the time-lag between
star formation and gas expulsion (ie. our simulations begin {\em
  at} gas expulsion).

If our interpretation is correct, the young ($<50$~Myr) clusters plotted on
Fig.~\ref{fig:clusters} formed with a wide range of eSFEs, somewhere
between 20 and 100~per cent.  We reiterate that the eSFE is not necessarily
a raw measure of the star formation efficiency -- rather it measures the
degree to which the cluster is out-of-equilibrium after gas
expulsion (see \S~2.1).

As clusters age they are either destroyed (eSFE$<30$ per cent), or relax into
(a new) virial equilibrium.  Therefore clusters older than $\sim 50$
-- $100$~Myr are expected to lie close to the canonical DMtL line as
is found in Fig.~\ref{fig:clusters}.  In some cases, even older
clusters may become perturbed (by tidal or impulsive interactions,
e.g. Gieles et al.~2006)
pushing them out-of-equilibrium causing them to move away from the 
canonical DMtL line, however we would expect this to be fairly rare.

We note that M82-F is the one cluster that does not fit into this
picture.  It lies significantly above the canonical line which would
suggest that it has an unusual top-heavy IMF (Smith \& Gallagher
2001), is contracting, or possibly that its age estimate is incorrect
(see BG06).  The two LMC clusters (NGC~1850 and NGC~2157) with ages 
of $\sim 30-40$~Myr in the middle of Fig.~\ref{fig:clusters} are also
rather unusual as their eSFE from their position in the fig. is
estimated to be $\sim 35$~per cent -- right on the border-line between
survival and destruction\footnote{Goodwin (1997b) estimates the eSFE
  of NGC~2157 to be 45 per cent, not too dissimilar to the 35 per cent
  suggested here.}.

\subsection{Infant mortality}

It is known that many young clusters {\em must} be destroyed.  Observations
of the age distribution of clusters in many galaxies show that the
number of clusters in a given (linear) age bin decreases significantly
with age.  In particular, very young ($< 20$~Myr) clusters are
significantly over-abundant.  Unless the cluster formation rate in
galaxies is significantly higher now than in the past
then many young clusters cannot survive into old age and suffer a high 
`infant mortality' (e.g. Lada \& Lada~2003, Bastian et al.~2005, Fall
et al.~2005).  Gas expulsion would seem an extremely good
candidate for the mechanism that causes (at least a significant 
fraction of) infant mortality.

Clusters with eSFEs below $\sim 30$~per cent cannot survive gas expulsion.
They are unable to find a new equilibrium and are destroyed (see also 
Goodwin 1997a,b; Kroupa \& Boily 2002; Boily \& Kroupa 2003a,b).  Such
destruction occurs in $10$ -- $30$~Myr (see
Fig.~~\ref{fig:massloss}).  The destructive expansion of low eSFE 
clusters is sufficiently rapid that clusters will likely 
become completely unobservable within a few 10s~Myr, as 
an increase in radius by a factor of 3 will cause the surface
brightness to decline by $\sim~2.5$~magnitudes (see Fall et al.~2005).

Examination of Fig.~\ref{fig:clusters} shows that of the 12 clusters
younger than $50$~Myr at least 7 of them had eSFEs of $< 30$~per cent
and so would be expected to be destroyed before they reach
$100$~Myrs.  This provides a lower limit on the infant mortality
in this sample of $\sim 50$~per cent.  However, the biases of
this sample are impossible to quantify as we rely on those clusters
for which there are dynamical mass estimates.  These clusters tend
to be bright which may suggest a bias towards high eSFE clusters
if there is any trend of eSFE with mass (e.g. Elmegreen \& Efremov
1997), but an examination of the data shows no significant trends (ie. eSFE
with luminosity or velocity dispersion).  We note however, that the
agreement between this estimate of the infant mortality rate agrees
well with other estimates, suggesting that the sample may not be 
very biased.

\subsection{Infant weight-loss}

As shown in Fig.~\ref{fig:massloss} even clusters that can survive gas
expulsion may lose a significant fraction of their stellar mass within 
the first $\sim40$~Myr.  Thus the star cluster mass function (CMF) 
may change drastically even over timescales as short as
this if the eSFE and cluster mass are correlated (see also 
Kroupa \& Boily~2002), which may be expected from cluster 
formation models (Elmegreen \& Efremov~1997).

If studies are restricted to the youngest clusters ($<10$~Myr) they
will measure the {\em initial} CMF with which clusters form.  If
clusters with a range of ages are included, then the effects of rapid
infant weight loss, early destruction through gas expulsion, and
slower destruction/weight loss by stellar evolution and/or tidal
fields must be accounted for.  Models of a full cluster population, 
such as those presented in Gieles et al. (2005), which fit cluster 
parameters in age and mass space simultaneously, are a promising 
way to measure whether there is a mass dependence in the (e)SFE.

We note that, even though a bound core may remain when the eSFE
is $>~30$~per cent, moderate eSFE clusters will become significantly
more vunerable to destruction.  Other effects which are not taken 
into account in this study (e.g. stellar evolution or tidal 
disruption) may be enough to dissolve the cluster completely.  Of 
particular importance may be interactions with GMCs (e.g. Gieles et 
al.~2006), as young clusters form in gas rich environments.

Thus we see that rapid gas loss and subsequent effects are 
an extremely rapid and efficient way to put newly formed stars into
the field.   There are also galaxy-scale implications of the effect of rapid
gas expulsion in young clusters (see Kroupa~2003).  Since star clusters
are born in a compact state, the velocity dispersion of the stars within the
clusters can be rather large as there is enough gravitational
potential energy provided by the remaining gas and other stars to bind
the cluster.  However, if the star-formation efficiency is low and
the residual gas is removed on time-scales shorter than a
crossing time, many stars  will be thrown into the field with large
velocities.  It has been suggested that this effect can alter the
morphology of galaxies and explain the age-velocity-dispersion
relation observed in the Galaxy (Kroupa~2003).  

\section{Conclusions}
\label{sec:conclusions}

We have simulated the evolution of massive young star clusters that
are far-out of virial equilibrium due to the expulsion of the residual
gas left-over after star formation.  Our results are in agreement with previous
simulations of this phase (e.g. Lada et al. 1984; Goodwin 1997a,b;
Geyer \& Burkert 2001; Kroupa et al. 2001; Kroupa \& Boily 
2002; Boily \& Kroupa 2003a,b).

We parameterise the initial state of a cluster with an effective star
formation efficiency (eSFE), $\epsilon$, where the initial virial
ratio of a cluster is $Q = 1/2\epsilon$ (we define virial equilibrium
to be $Q=0.5$).  We explore clusters with eSFEs
between 10 and 60 per cent.

Clusters expand after gas expulsion in an attempt to reach a new
equilibrium.  If eSFE $\leq 30$ per cent then the clusters are destroyed
within a few 10s~of Myr, for higher eSFEs a bound core remains, but there
may still be a significant degree of (stellar) mass loss and the final
cluster may only be a small fraction of its initial mass.  Those stars
that escape cause an excess of light at large radii that is observed
in many young clusters (BG06).

During its expansion a cluster is not in virial equilibrium.  Its
velocity dispersion retains a memory of the initial (gas plus stars)
mass of the cluster and so dynamical mass estimates -- which assume
virial equilibrium -- may be wrong by a very significant factor.

We also note that during the expansion phase cluster parameters, such
as the core radius, may change significantly on a short timescale.  In
particular they may not change in a linear fashion.  The core radii
of intermediate eSFE clusters measured when escaping stars are still
physically associated with the cluster may be significantly greater
than the `final' (relaxed) core radii (see also Goodwin 1997b).
Many properties are {\em
  instantaneous} and it may be difficult to extrapolate to future
values without knowing the initial conditions.  This includes the mass
of clusters which can change significantly as stars are lost (see also
Kroupa \& Boily 2002).

We compare the dynamical (assuming virial equilibrium) and luminous 
(assuming a standard Kroupa IMF) masses for a large sample of young
clusters (limited by the number of clusters for which reliable values
for both parameters are available).  As noted by Bastian et
al. (2006), young ($<50$~Myr) clusters tend to have too much dynamical mass for
their luminosity, whilst older clusters have dynamical masses that
match their expected luminosity for a Kroupa IMF.  

We suggest that the discrepancy between the dynamical and luminous
masses for young clusters is due to the incorrect assumption of
virial equilibrium.  We assume that young clusters -- the same as old
clusters -- have a standard IMF and use the discrepancy between the
mass estimates to calculate their eSFEs which appear to fall between
$20$ and $\sim 100$ per cent (an upper limit is very difficult to
establish as for eSFEs$>60$ per cent the effect of gas expulsion is
difficult to observe).  This wide range of eSFEs may be due to a real
difference in the star formation efficiencies of different clusters,
or to differences in the initial equilibrium between the stars and gas
(or, most likely, a combination of the two).

At least $50$ per cent of the young clusters have eSFEs $<30$ -- $40$
per cent and so we expect the infant mortality rate in this sample to
be {\em at least} $50$ per cent (as we ignore further destructive
effects such as stellar evolutionary mass loss and tidal fields).
Further, there is an observational bias that means that low surface 
brightness -- hence more extended, or more expanded -- clusters which
are likely to be destroyed are not observed.

In summary, young clusters are most likely to be re-virialising after
gas expulsion.  This means that (a) dynamical mass estimates may be
significantly wrong, and (b) measured parameters (including mass) are
instantaneous values and may not reflect the `final' values.  Assuming
that luminosity-derived masses for young clusters are correct allows
us to determine the effective star formation efficiencies of young
clusters with dynamical mass estimates and we find that at least $50$
per cent of young clusters in our sample are unlikely to survive for
more than a few 10s~Myr.  The true infant mortality rate is probably
much higher than this.

\section*{Acknowledgments}

SPG is supported by a UK Astrophysical Fluids Facility (UKAFF)
Fellowship.  The GRAPE-5A used for the simulations was purchased on
PPARC grant PPA/G/S/1998/00642.

\bsp
\label{lastpage}
\end{document}